\documentclass[12pt,preprint]{aastex}

\shorttitle{}
\shortauthors{de Mello et al.}

\begin{document}


\title{Star-formation in the HI bridge between M81 and M82\altaffilmark{1}}


\author{D. F. de Mello\altaffilmark{2,3,4} }
\author{L. J. Smith\altaffilmark{5,6,7}, E. Sabbi\altaffilmark{5}, J.S. Gallagher\altaffilmark{8}, M. Mountain\altaffilmark{5} \& D.R.  Harbeck\altaffilmark{8}}

\altaffiltext{1}{Based on observations with the NASA/ESA {\it Hubble
    Space Telescope}, obtained from the Data Archive at the Space Telescope Science
    Institute, which is operated by AURA, Inc., under NASA contract
    NAS5--26555. These observations are associated with program \#10915.} 
\altaffiltext{2}{Observational Cosmology Laboratory, Code 665, Goddard Space Flight Center, Greenbelt, MD
20771}
\altaffiltext{3}{Catholic University of America Washington, DC 20064}

\altaffiltext{4}{Johns Hopkins University, Baltimore, MD 21218}

\altaffiltext{5}{Space Telescope Science Institute, 3700 San Martin Dr., Baltimore, MD 21218}

\altaffiltext{6}{ESA Space Telescope Operations Division}

\altaffiltext{7}{Department of Physics and Astronomy, University College London, Gower Street, London WC1E
6BT, UK}

\altaffiltext{8}{Department of Astronomy, University of Wisconsin-Madison, 475 North Charter Street,
Madison, WI 53706}


\begin{abstract}
We present multi-wavelength observations of stellar features in the H\,{\sc i} tidal bridge connecting M81 and M82 in 
the region called Arp's Loop. We identify eight young star-forming regions from Galaxy Evolution Explorer ultraviolet observations. 
Four of these objects are also detected at H$\alpha$.  We determine the basic star formation history of Arp's Loop using F475W and F814W 
images obtained with the Advanced Camera for Surveys on board the Hubble Space Telescope. We find both a young ($<$ 10 Myr) and an old ($>1$ Gyr) 
stellar population with a similar spatial distribution and a metallicity $Z\sim0.004$.
We suggest that the old stellar population 
was formed in the stellar disk of M82 and/or M81 and ejected into the intergalactic medium during a 
tidal passage ($\sim$ 200--300 Myr ago), whereas the young UV-bright stars have formed in the tidal debris. The UV luminosities of the eight 
objects are modest and typical of small clusters or OB associations. 
The tidal bridge between M81--M82 therefore appears to be intermediate between the very low levels  
of star formation seen in the Magellanic bridge and actively star-forming tidal tails associated with major galaxy mergers.
\end{abstract}





\section{Introduction}

The M81 group contains three major closely interacting galaxies: M81, M82 and NGC 3077. Studies of this triple system at radio wavelengths have shown an extremely disturbed H{\sc i} distribution with tidal bridges connecting the three galaxies (e.g. Gottesman \& Weliachew 1975;  
van der Hulst 1979; Appleton, Davies \& Stephenson 1981; Yun, Ho \& Lo 1994). Within the tidal bridges, H{\sc i} knots with optical counterparts are seen. The most prominent of these to the east/north-east of M81 are Holmberg IX 
and Arp's Loop or A0952$+$69 (Arp 1965). It has been suggested (e.g. Boyce et al. 2001) that these are forming tidal dwarf galaxies. Numerical simulations of the system by Yun (1999) can successfully reproduce the 
the H{\sc i} tidal debris if the closest approaches of M82 and NGC 3077 to M81 took place 220 and 280 Myr ago respectively. Arp's loop, the topic of this paper, occurs at the intersection of the three tidal streamers in 
the simulation of Yun (1999) who gives an H{\sc i} mass of $2.6 \times 10^8$ M$_\odot$. Sun et al. (2005) present very deep wide-field optical continuum images of the M81/M82 galaxy group, and show that groups of young stars are distributed across the entire region between the two galaxies. 

Although tidal tails usually have blue optical colors (e.g. Schombert et al. 1990), it is not clear whether these colors are due to tidally-disrupted stellar disk material or active star formation in the tails.
Recently, ultraviolet (UV) imaging obtained with the Galaxy Evolution Explorer (GALEX) satellite has shown that regions of bright UV emission, containing recently formed stars, correlate well with regions of high H{\sc i} column density in tidal tails (e.g. Hibbard et al. 2005; Neff et al. 2005). These findings suggest that the combination of UV and H{\sc i} data provides a powerful technique for identifying and studying star formation in the intergalactic medium (IGM) between interacting galaxies. This topic is of importance for earlier epochs in the universe because galaxy interactions, and thus tidal tails were more common, leading to a higher star formation rate in the IGM. This may well have implications for the enrichment of the IGM; Ryan-Weber et al. (2004) have shown that stars forming in the IGM offer a viable alternative to galactic wind enrichment on large scales.

We have thus started a study of star formation in the IGM of the M81 galaxy group with the overall aim of elucidating the importance of this mode of
star formation. We present results for Arp's loop in this paper and for 
Holmberg IX in Sabbi et al. (2007). Arp's loop is located 17 arcmin north-east of M81 and its composition was first discussed by Efremov et al. (1986). They resolved the ring-like structure into blue diffuse and star-like objects 
which they interpreted as young star clusters and associations. Makarova et al. (2002) investigated the stellar population of Arp's loop using photometry obtained with the Wide Field and Planetary Camera 2 (WFPC2) on board the 
Hubble Space Telescope (HST). They were unable to clarify the evolutionary nature of Arp's loop because of photometric limitations and WFPC2's small field of view.  
They find evidence for: enhanced star formation between 20 and 200 Myr ago, coincident with the likely epoch of tidal interactions in the M81 group; intermediate age stars 
which were either formed prior to the interaction in the Arp's loop or were torn out of the massive galaxies during the interaction; 
and also faint stars with ages $\ge$ 8 Gyr but which were too close to the detection limit for detailed analysis.
In this paper, we use multi-wavelength observations, including deep photometry obtained with the Advanced Camera for Surveys (ACS), to resolve 
the stellar population, and thus determine the evolutionary status of Arp's loop.

\section {The Data}

\subsection{GALEX, ACS, H{\sc i} and WIYN}

The M81 group was observed with the Galaxy Evolution Explorer (GALEX) mission in the far 
(FUV: $\lambda$$_{\rm eff}$=1528\AA; $\Delta$$\lambda$$_{FUV}$=269\AA)
and near ultraviolet (NUV: $\lambda$$_{\rm eff}$=2271\AA; $\Delta$$\lambda$$_{NUV}$=616\AA). We obtained the data from the Multimission Archive at STScI (MAST) as
part of the team General Releases. The
GALEX field of view is 1$^{\circ}$.28 and 1$^{\circ}$.24 in the FUV and NUV bands, and the pixel scale is 1.5$''$/pixel. The exposure times were 3093s for each band. 

Fluxes were calculated using the equations of Morrissey et al. (2005): m$_{\lambda}=-2.5$ log[F$_{\lambda}$/a$_{\lambda}$] + b$_{\lambda}$, 
where a$_{FUV}$ = 1.4 $\times$ 10$^{-15}$ erg s$^{-1}$ cm$^{-2}$ \AA$^{-1}$,  
a$_{NUV}$=2.06$\times$ 10$^{-16}$ erg s$^{-1}$ cm$^{-2}$ \AA$^{-1}$, b$_{FUV}$=18.82 and b$_{NUV}$=20.08 for FUV and NUV, respectively. Fluxes were multiplied
by the effective filter bandpass ($\Delta$$\lambda$$_{FUV}$=269\AA\, and $\Delta$$\lambda$$_{NUV}$=616\AA) to give units of erg s$^{-1}$ cm$^{-2}$ and  
luminosities were calculated for a distance of 3.6~Mpc (Freedman et al. 1994).

\begin{figure}
\epsscale{.80}
\plotone{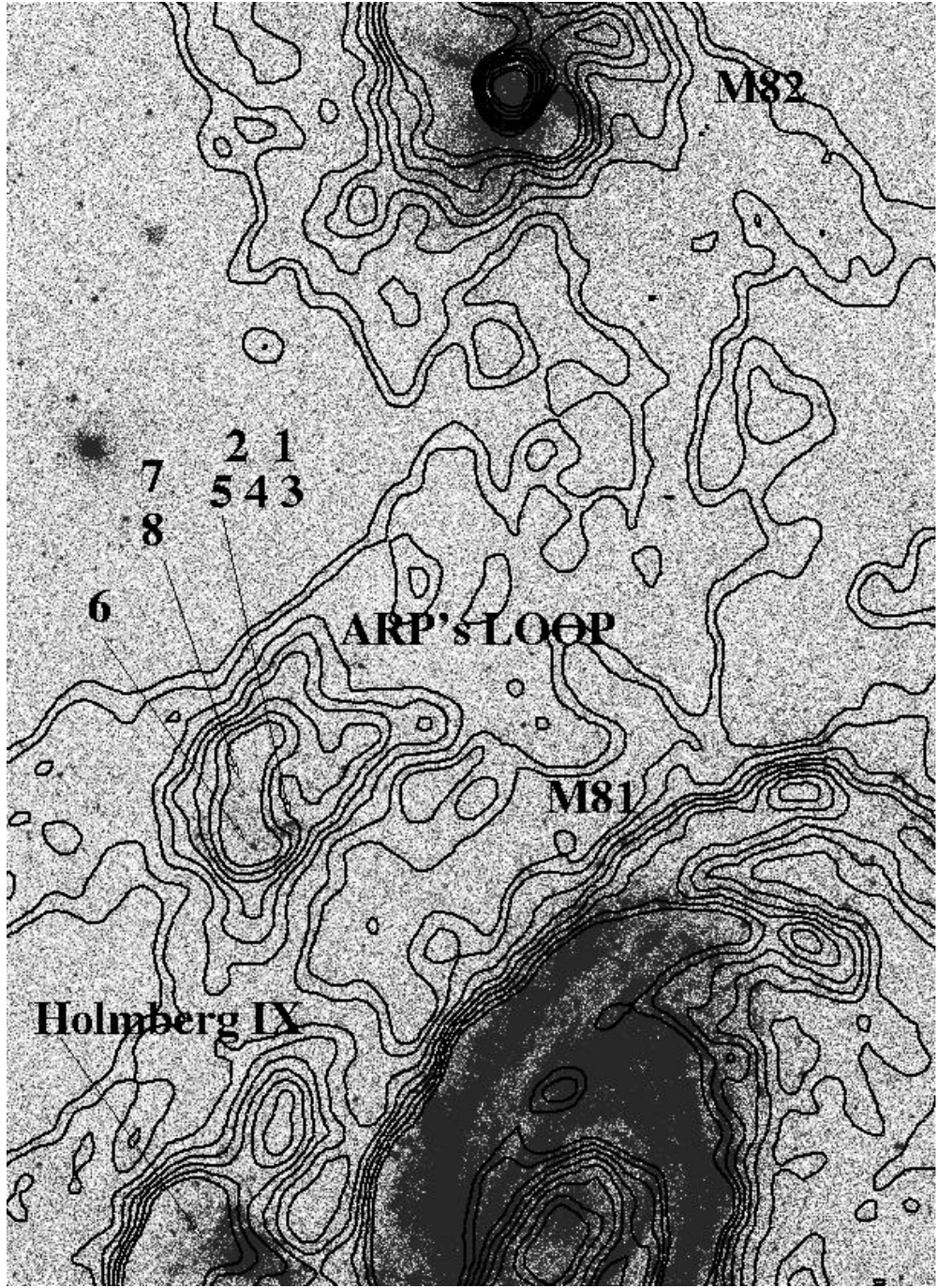}
\caption{GALEX FUV image of the M81 and M82 region where FUV objects are detected. 
The contours in black are HI contours from Yun et al. (1994). FUV objects are numbered 1-8 and the general location of the region 
called Arp's loop is marked. North is to the top and
East to the left, M81 is shown partially in the bottom of the image and M82 in the top of the image, Holmberg~IX is to the east of M81, 
the cutout is 31.1$'$ $\times$ 42.6$'$. H{\sc i} contours adapted from 
Yun, Ho \& Lo's Fig.~1 by permission from Macmillan Publishers Ltd: Nature, copyright (1994).\label{m81m82hi}}
\end{figure}

\begin{figure}
\epsscale{1.0}
\plotone{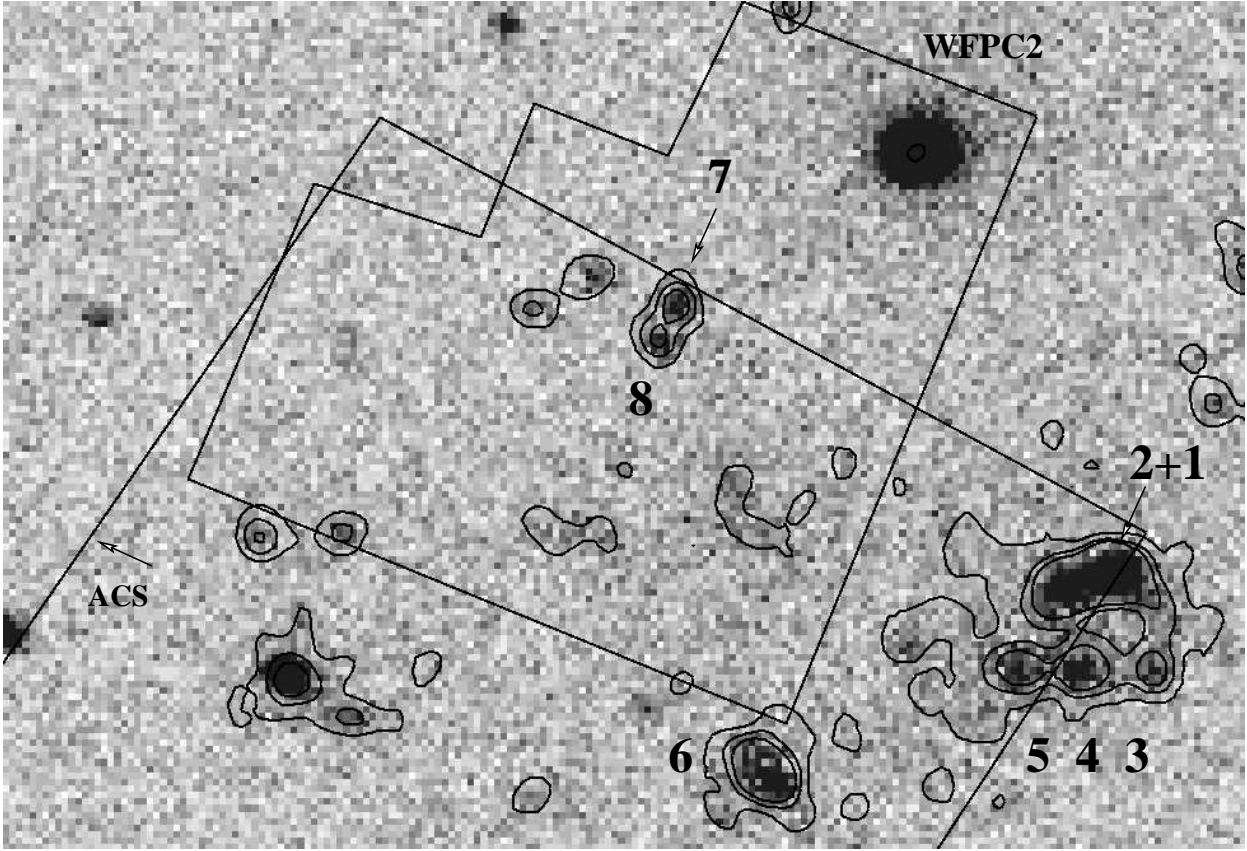}
\caption{GALEX NUV image with FUV contours in black (contour levels are 0,
0.0006, 0.009 and 0.001 counts).
The blue blobs numbered 1-8 in Fig.~\ref{m81m82hi}, the ACS footprint, and
WFPC2 region used in Makarova et al. (2002) are marked. The cutout is
4.8$'$ $\times$ 3.2$'$. North is to the top and East to the left.
\label{nuv_feet}}
\end{figure}

\begin{figure}
\epsscale{1.0}
\plotone{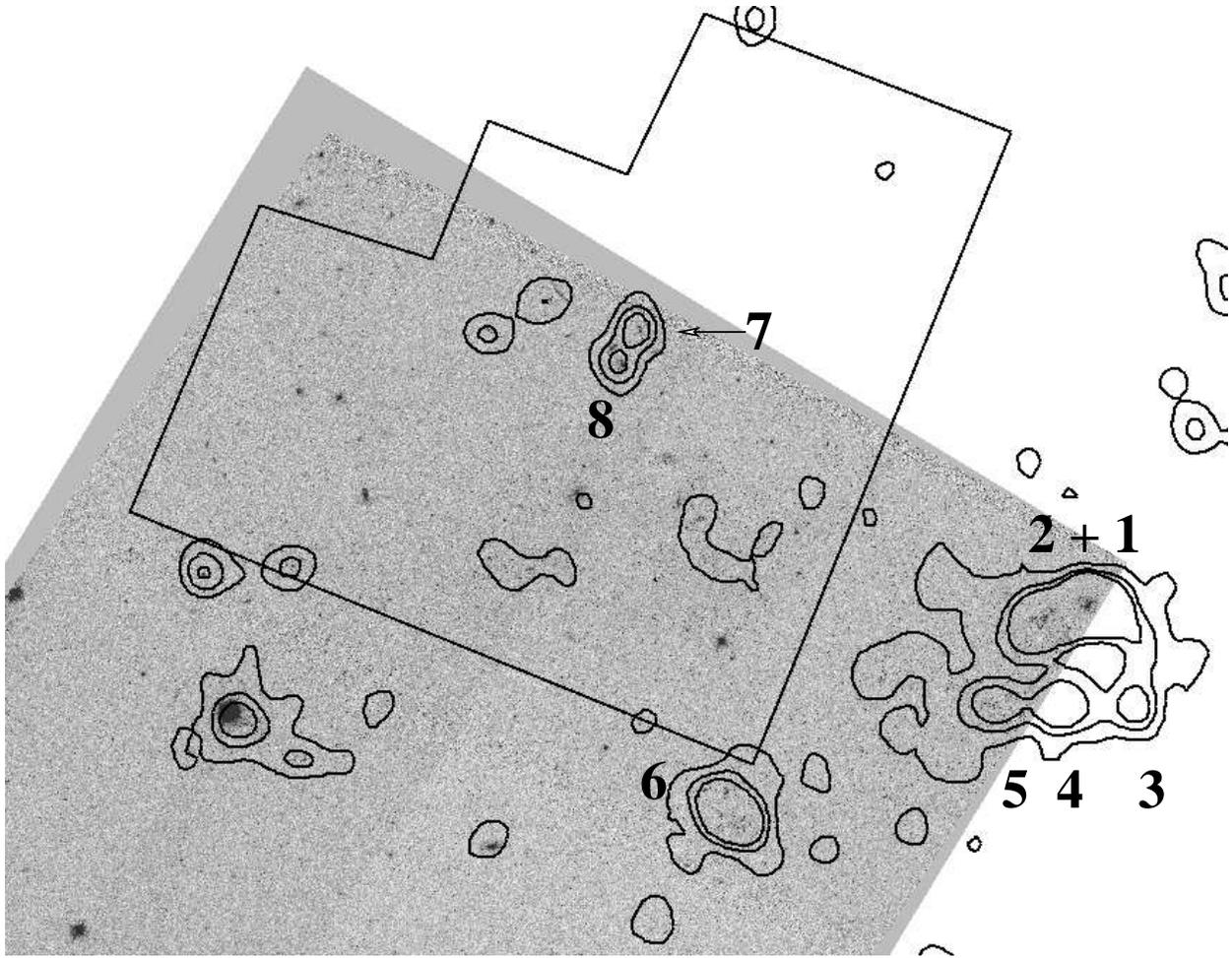}
\caption{ACS/F475W image of the region where FUV objects are detected (4.5$'$ $\times$ 3.3$'$). GALEX FUV contours are shown in black. 
WFPC2 region used in Makarova et al. (2002) is marked. North is to the top and
East to the left.
\label{acs_feet}}
\end{figure}

\begin{figure}
\plotone{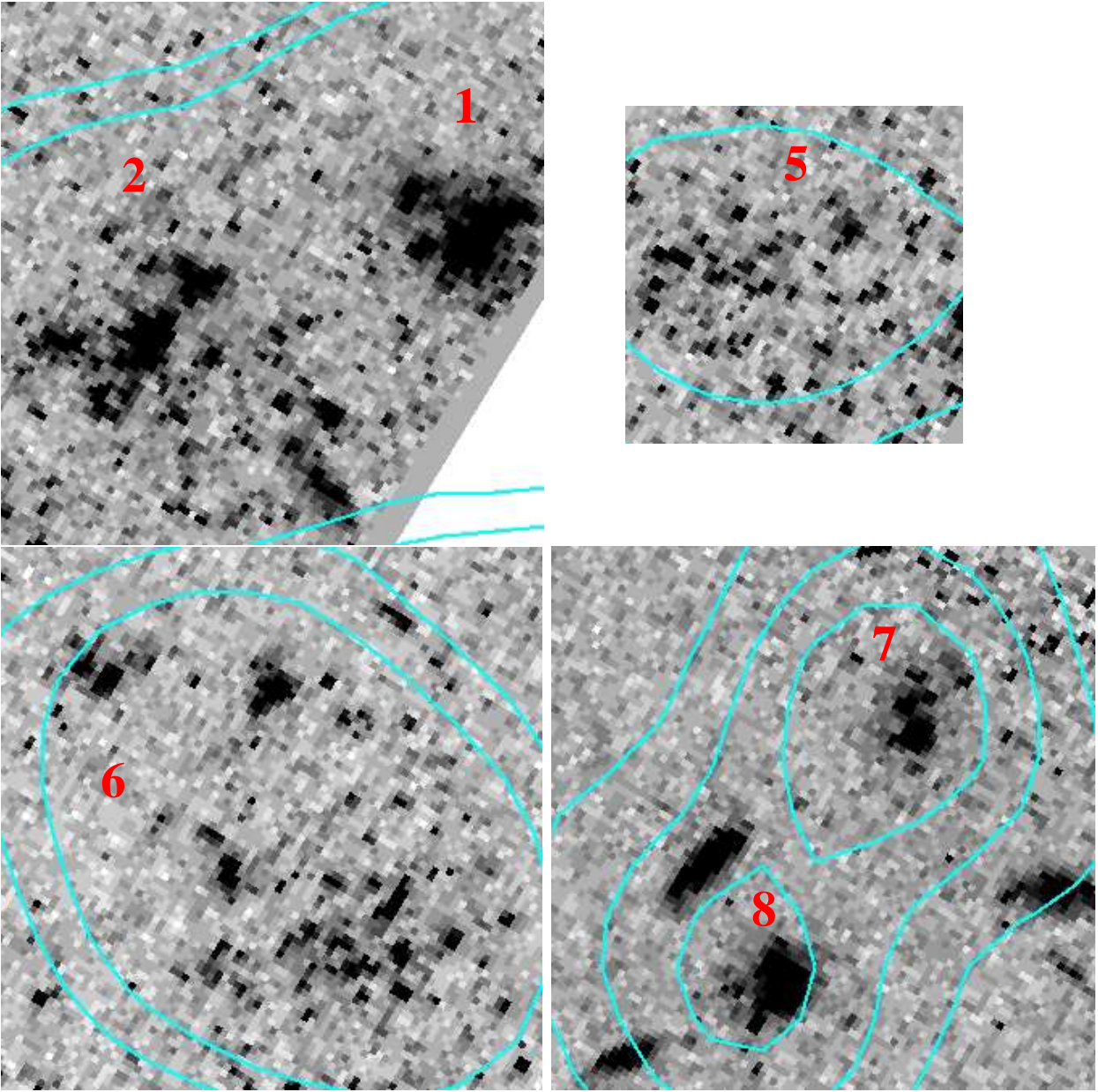}
\caption{ACS/F475W image cutouts. Objects~\#1 and 2 are shown within a 16$''$ $\times$ 16$''$ box, objects\#3 and 4 are outside the ACS pointing, object~\#5 is shown 
within a 10$''$ $\times$ 10$''$ box, object \#6 is shown within a 16$''$ $\times$ 16$''$ box, objects~\#7 and 8 are shown within a 
16$''$ $\times$ 16$''$ box. GALEX FUV contours are shown in cyan. North is to the top and
East to the left.
\label{blobszoom}}
\end{figure}

Figure \ref{m81m82hi} shows the distribution of the atomic hydrogen in the M81 group of galaxies (Yun, Ho \& Lo
1994) on top of the GALEX FUV image. We have identified 8 FUV objects (Fig.~\ref{nuv_feet}, Table 1) inside the H{\sc i} cloud located on the northeast side of 
the bridge between M81 and M82 and north of Holmberg~IX. This region, called concentration~II by Yun et al., is the brightest 
H{\sc i} complex in the tidal bridge. It includes Arp's loop, and may not be a single cohesive feature, containing tidally stripped gas from M81 and 
M82. Makarova et al. (2002) found 250 stars in the HST/WFPC2 F606W/F814W images of Arp's loop. However, the WFPC2 observations were centered on the H{\sc i} 
concentration and cover only objects \#7 and \#8, missing 
the other FUV objects. We searched the HST archive at MAST and found new pointings of the same area but taken with the  Wide Field Channel (WFC) of ACS using the F475W and F814W filters and covering a larger field of view (GO 10915: PI Dalcanton). Fig.~\ref{acs_feet} shows the ACS F475W image where we have plotted the WFPC2 footprint, the 
 GALEX contours, and marked the 8 FUV objects. The high resolution of ACS (0.05$''$) resolves the FUV objects into hundreds of starlike objects as can be seen in 
 Fig.~\ref{blobszoom}. However,  
 the ACS pointing misses two of the FUV objects (\#3 and \#4) and part of another object (\#1). 
 
\begin{figure}
\epsscale{1.0}
\plotone{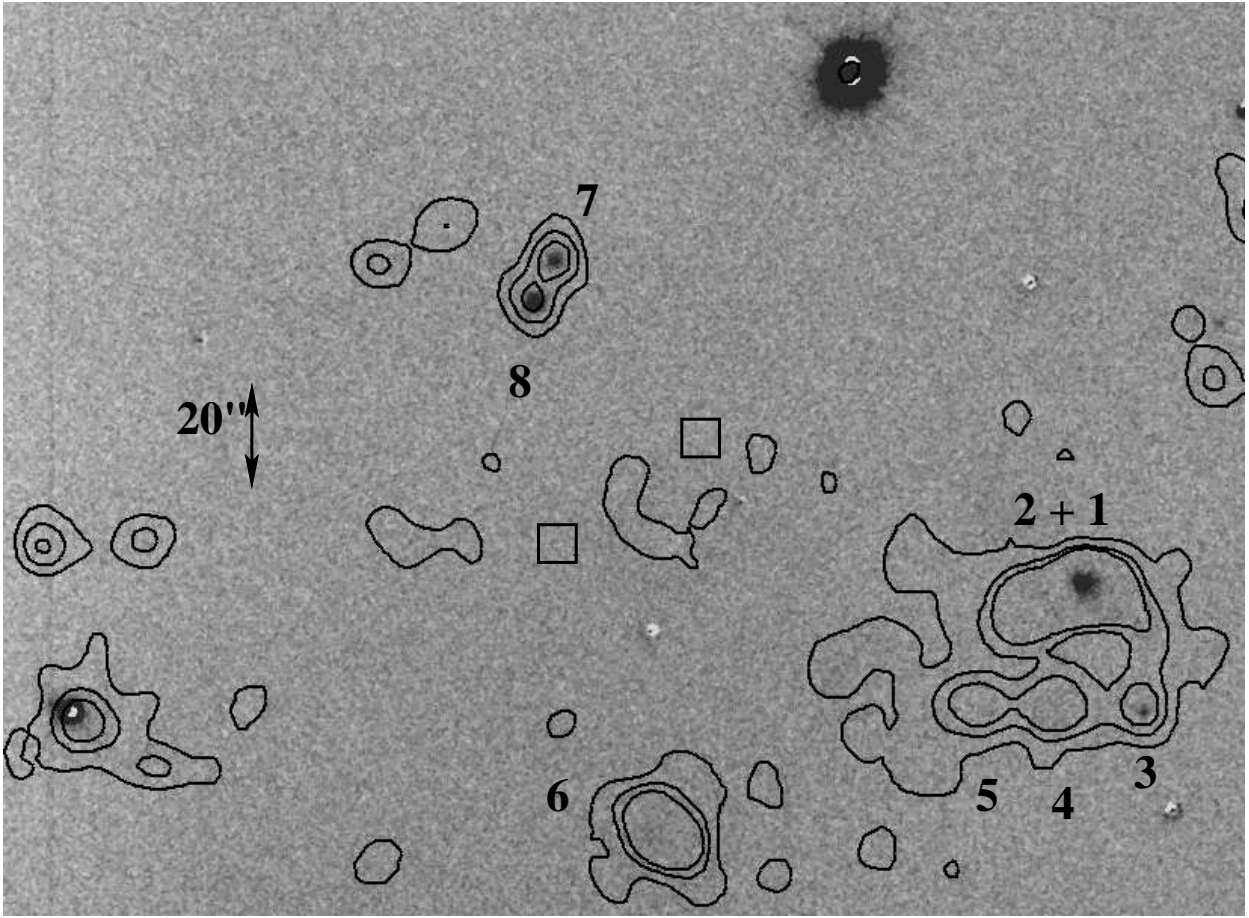}
\caption{WIYN 3.5m telescope H$\alpha$ continuum subtracted (smoothed with a 5$\times$ 5 median filter) image. The two boxes show the location of two objects detected by 
Karachentsev \& Kaisin (2003) but not present in our data. GALEX FUV contours are shown in black. North is to the top and
East to the left.
\label{wiyn}}
\end{figure}

We have also obtained an H$\alpha$ image of Arp's loop with the OPTIC  
imager at the WIYN{\footnote {The WIYN Observatory is a joint facility of the University of
    Wisconsin-Madison, Indiana University, Yale University, and the
    National Optical Astronomy Observatories.}}
 3.5m telescope. The observations were taken with  
WIYN filter \# 13 ($\Delta$$\lambda$ =  47 \AA) and a standard Johnson $R$
filter for continuum subtraction. OPTIC is a  
mosaic of two 2kx4k orthorgonal transfer CCDs, resulting in a field  
of view of 9.5$'$$\times$9.5$'$. In Fig.~\ref{wiyn} we present only an image of chip \#2 which was bias-corrected and flat-fielded using IRAF{\footnote{IRAF is distributed by the National Optical Astronomy Observatory, which is operated by AURA, Inc., under cooperative agreement with the National Science Foundation.}}
The two images taken in each filter were combined with cosmic-ray rejection,  
and the resulting scaled $R$-band image was subtracted from the  
combined H$\alpha$ image to remove the continuum signal. Our data confirm Karachentsev \& Kaisin's (2007)  
 detection of objects \#1, \#8, \#7, and \#3, with \#1 being the brightest and \#3 the faintest object (see fig.1 in Karachentsev \& Kaisin). 
 As shown in Fig.~\ref{wiyn}, although we have detected the faintest object, 
 we have not detected two of their H$\alpha$ sources which are marked as squares in Fig.~\ref{wiyn}. 
 This could be due to the different bandwidths, since Karachentsev \& Kaisin used 
 a broader filter ($\Delta$$\lambda$ = 75 \AA) than ours. H$\alpha$ detections of the FUV sources  
 suggest that these are small H{\sc ii} regions where hot, blue OB stars are emitting ultraviolet radiation. 

\section{Color-Magnitude Diagram}

In order to investigate the stellar content of the FUV objects, we have performed photometry on the two ACS/WFC images. 
The total exposure times for the  
fully reduced and multidrizzled images are 2250 s in the F475W ($\sim B$) band and  
2265 s in the F814W ($\sim I$) band respectively. Each image covers an area of $200 
\arcsec \times 200\arcsec$, which  corresponds to $3.5\times 3.5\,  
{\rm kpc}$ for the distance of M81.
The images were reduced using the standard STScI ACS pipeline {\sc calacs}  
(Pavlovsky et al. 2004) to remove bias, dark current and flat--field  
response signatures. The dithered images were combined using the 
MultiDrizzle package, which removes cosmic rays, and corrects the  
images for geometric distortion.

Photometry was performed on the drizzled images using the DAOPHOT package  
within IRAF. Stars were automatically detected with DAOFIND in the F475W  
band, with the detection threshold set at 4 $\sigma$ above the local  
background level; the detections were then forced on the $I$ band image.  
The instrumental magnitude of each object in each filter was  
estimated via a PSF--fitting technique. A spatially variable PSF was  
computed for both the F475W and F814W images, using $\sim 180$ isolated stars  
in different positions on the images.

Selection criteria based on the shape of the objects have been  
applied to the photometric catalog to remove as many spurious detections, 
blended and/or extended objects as possible. We have  
considered the DAOPHOT ``sharpness'' parameter, which sets the intrinsic angular size  
of the objects, and selected only those with $-0.5 < {\rm sharpness} <  
0.5$ in both filters. These sharpness values allow us to reject spurious and extended objects without  
eliminating the bright stars. We emphasize that the FUV objects are  
only partially resolved in the ACS data: one WFC/ACS pixel corresponds  
to 0.05$''$, which is $\sim$1.5 pc at the distance of M81. 
Our final photometry is calibrated on the ACS {\sc vegamag} photometric  
system, with zero points adopted from Sirianni et al. (2005).

We have also measured the optical integrated flux at $V$ of the FUV objects detected by GALEX 
by averaging the F475W and F814W images, and using the {\sc polyphot} routine in  IRAF with the GALEX FUV contours, 
to determine $V$ magnitudes in the AB system. Although the $V$ magnitudes were measured in a pseudo $V$ image, 
we included it in Table 1 since it is a standard magnitude provided in the literature.

\begin{figure}
\plotone{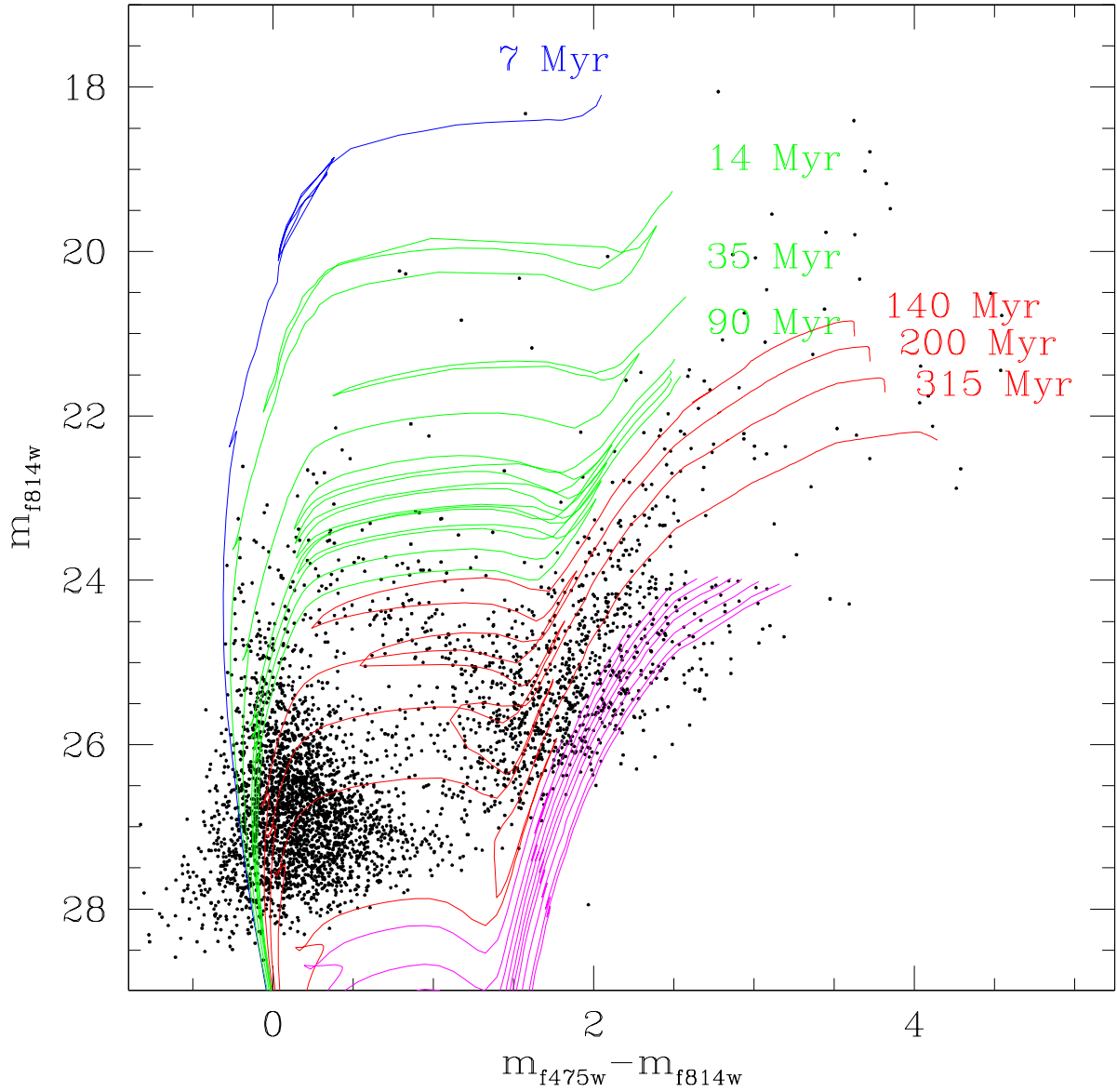}
\caption{Color-magnitude diagram, $F814W$ versus $F475W-F814W$ ($\sim I versus B--I$), 
of all stars measured in the ACS field of view. Superimposed on the CMD are Padova isochrones 
(Fagotto et  al. 1994a,b), with a metallicity Z=0.004 and ages 7 Myr to 10 Gyr.
\label{cmd}}
\end{figure}

Figure~\ref{cmd} shows the $F814W$ versus $F475W-F814W$ ($\sim B-I$) 
color--magnitude diagram (CMD) for  
all stars measured in the ACS field of view. The CMD shows a  
young and blue main sequence (MS) evolutionary phase, located at $F475W-F814W \sim -0.1$, with the brightest stars at $I\simeq 22.5$. A scarcely  
populated red plume is visible at $B-I>1.5$ and extends up to $I 
\sim21.0$. It is populated by red supergiants (RSGs) at the brighter  
magnitudes, and asymptotic giant branch stars (AGBs) at fainter  
luminosities. One of the most interesting features in the CMD is the diagonal  
strip of stars extending through $F475W-F814W\simeq 0.5$, $F814W\simeq 23$ and   
$F475W-F814W\simeq 1.5$, $F814W\simeq 27$. This strip is populated by stars at  
the hot edge of the helium core-burning phase. The concentration of stars in the range of colors $2<F475W-F814W<3$ 
and magnitudes $24<F814W < 27$ corresponds to low mass ($M<2.2\,  
{\rm M}_\odot$), older (age $> 1\, {\rm Gyr}$) stars in the red giant  
branch (RGB) evolutionary phase.

We used Padova isochrones for metallicity $Z=0.001, Z=0.004$ and $Z=0.019$
(Fagotto et al. 1994a,b) to enable us to interpret the features visible in the CMD. 
Superimposed on Fig. 7 is the set of Padova isochrones for a metallicity $Z 
=0.004$. We obtain this value for $Z$ from the following considerations. For the 
old population, we can rule out metallicities as high as solar from the shape and 
broadening of the RGB. For the same reasons, we find that $Z$ is unlikely to be $< 
0.004$. However we cannot exclude a slightly higher metallicity.
The color and morphology of the blue edge of the blue loop evolutionary phase 
are sensitive to $Z$ and age. We find that the colors of this structure can only be
reproduced for $Z \approx 0.004$, assuming that star formation began $\sim 300$ Myr ago in Arp's Loop and is continuing to the present. We cannot put strong constraints 
on $Z$ for the young stellar population because there are not enough stars, and 
some of these stars may be multiple.

\begin{figure}
\plotone{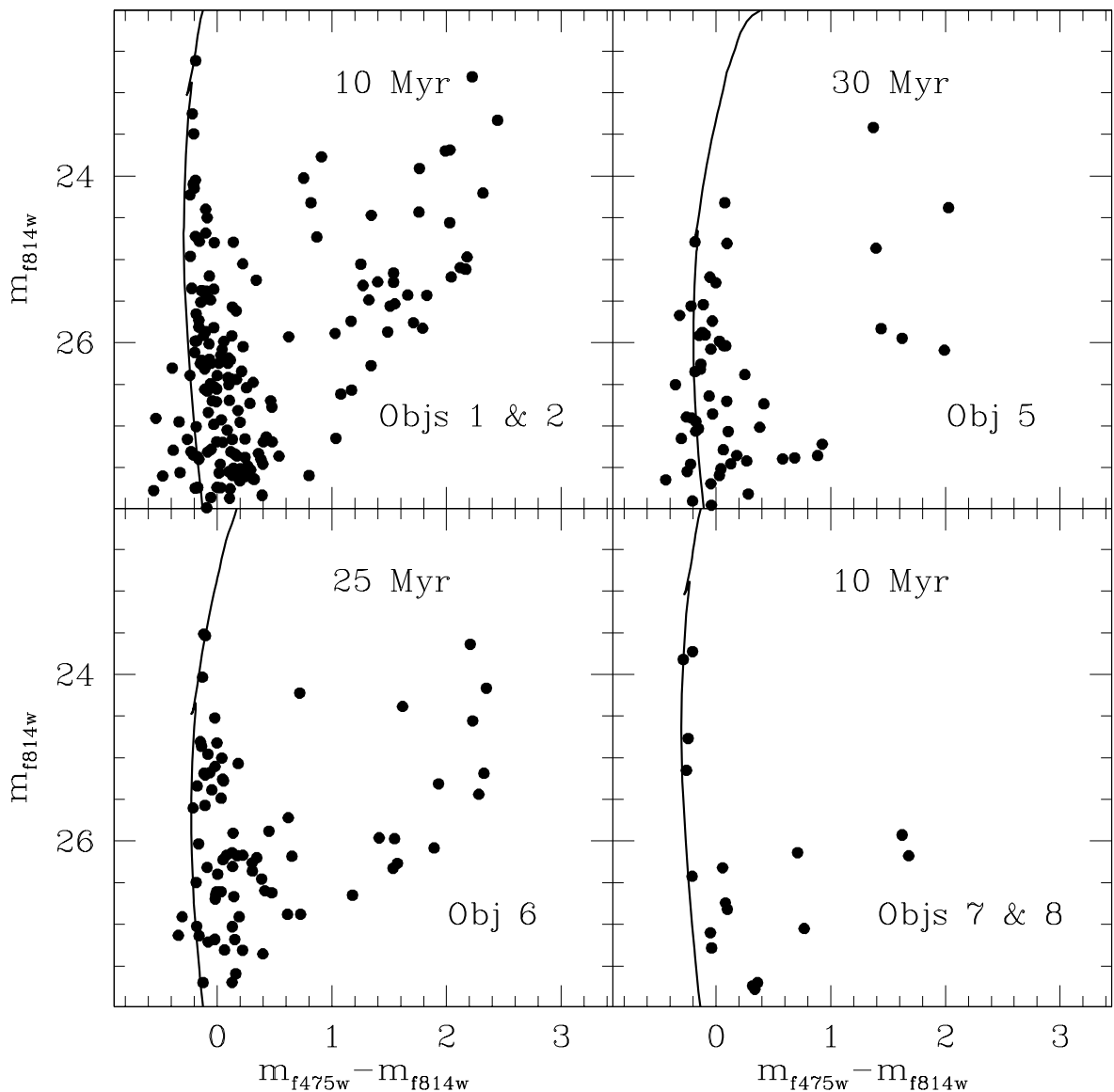}
\caption{Color-magnitude diagram,$F814W$ versus $F475W-F814W$ ($\sim I versus B--I$), of the FUV regions 1-8.  Superimposed on the CMD are Padova isochrones 
(Fagotto et  al. 1994a,b), with a metallicity Z=0.004 and ages 10 (regions \# 1$+$ 2, and \# 7$+$8) 25 (region \# 6) and 30  Myr (region \# 5).
\label{cmdblobs}}
\end{figure}

The youngest stars  are
concentrated where GALEX detected the FUV regions. We used GALEX  
isophotal contours to select stars within the FUV regions. CMDs  
corresponding to FUV objects \#1 \& \#2, \#5, \#6, and \#7 \& \#8 are  
shown in Fig.~\ref{cmdblobs}. A comparison with Padova isochrones indicates that 
the regions where the FUV objects are detected are mostly dominated by O and B stars 
with ages younger than 30 Myr. 

Individual O stars are difficult to distinguish in CMDs based on optical filters because of their well known color degeneracy at these wavelengths. We also have the additional problem that any O stars in Arp's Loop may be multiple objects (one WFC/ACS pixel corresponds  to  $\sim$1.5 pc).
To estimate the number of O stars in Arp's Loop, we can use both the GALEX FUV fluxes and the nebular H$\alpha$ luminosity of Karachentsev \& Kaisin (2007). We measure a total FUV flux of 
$2 \times 10^{39}$ erg s$^{-1}$ (Table 1) and, using standard numbers,  this is equivalent to $\sim$ 50 O8V stars. This is an upper limit because we have neglected the contribution of B stars to the FUV flux. Karachentsev \& Kaisin (2007) measure a total H$\alpha$ flux of
$3.5 \times 10^{-14}$ erg s$^{-1}$ cm$^{-2}$ giving $L$(H$\alpha) = 5.4 \times 10^{37}$ erg s$^{-1}$.
Thus, assuming one O8V star has a photon ionizing luminosity $Q_0$ of $3 \times 10^{48}$ s$^{-1}$ in the Lyman continuum (Smith, Norris \& Crowther 2002), the observed 
H$\alpha$ flux is equivalent to 13 O8V stars. 
These rough estimates will increase if there is any internal extinction. Arp's Loop is then equivalent to $\sim 5$ Orion nebulae, if we assume that the Orion Nebula has an H$\alpha$ luminosity of $10^{37}$ erg s$^{-1}$.

\begin{figure}
\plotone{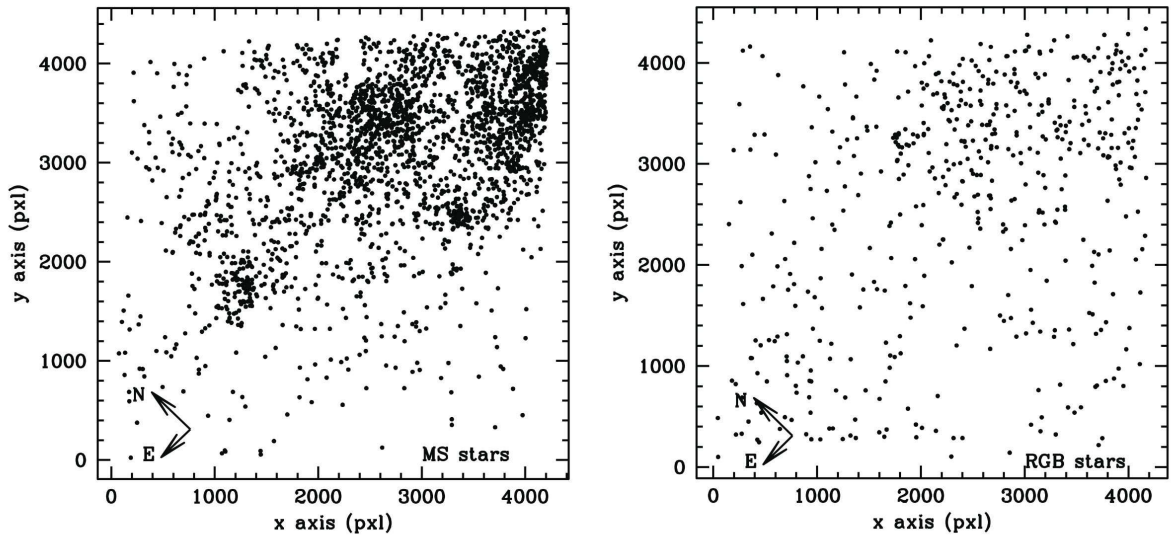}
\caption{Spatial distribution of main sequence stars (left) and RGB stars (right) in the ACS image. Orientation N-E is shown in each plot.
\label{spatial}}
\end{figure}

Fig.\ref{spatial} shows the spatial distribution of the main sequence and the RGB stars in the ACS image. It is clear that
stars of both types are located in the same general areas of the image with only two exceptions, at (1400,1600) and (3400, 3400).
These regions of the image have many MS stars and only a couple of RGBs are seen. The first region is not among the GALEX sources, it is 
close to the bright star on the south-east side of Fig.\ref{acs_feet} and the other region corresponds to object \#6.

In summary, we find both an old and a young stellar population in Arp's Loop with a similar spatial distribution and a metallicity $Z \sim 0.004$.
It is likely that the old stellar population 
was formed in the M82 and/or M81 galaxies and ejected into the intergalactic medium during a 
tidal passage ($\sim$ 200--300 Myr ago), whereas the young stars detected by GALEX have formed since the encounter in the tidal debris.

\section{Discussion}

\subsection{Star Formation in Tidal Debris}
Observational evidence is accumulating that star formation often
occurs in low gas density environments. Examples include the outer
disks of galaxies (e.g. Ferguson et al. 1998, Thilker et al. 2005)
and tidal tails (e.g. Saviane, Hibbard, \& Rich 2004, Hibbard et al.
2005). The nearest example is the H{\sc i} bridge between the
Magellanic Clouds (see Demers \& Battinelli 1998), located at a
distance of 50--60~kpc which allows very detailed investigations of
both its interstellar gas and stars. The bridge has a complex
structure in H{\sc i} with typical column densities of 10$^{20}$--10$^
{21}$~cm$^{-2}$ (Muller et al. 2003, 2004) and somewhat higher in smaller regions.  The Harris
(2007) optical survey confirms that stars along the Magellanic bridge
not only formed in place but also formed in material that contains
very few old red giant stars. Molecular line surveys (Mizuno et al.
2006) and follow-up searches for pre-main-sequence stars (Nishiyama et
al. 2007) demonstrate that despite overall low H{\sc i} column
densities, the conditions necessary for star formation and young
stellar candidates exist in some places along the western part of the bridge
that extends from the wing of the Small Magellanic Cloud.  The more
diffuse Magellanic stream with $N$(H{\sc i}) $ \lesssim 3$--$5\times 10^{20}$ cm$^{-2}$
(e.g. Br\"{u}ns et al. 2005), however, is not known to contain any
stars (Majewski et al. 1999).

At the other extreme, relatively intense star formation is detected
in FUV images of more massive gas rich tidal tails
(e.g. Hibbard et al. 2005 for the Antennae, and Neff et al. 2005 for 
NGC 7769/71, NGC 5713/19, Arp 295, and the NGC 520 system).
In these cases, large  H{\sc ii} regions, OB associations, and even 
super star clusters can form,
especially in situations where the H{\sc i} column densities are 
$N$(H{\sc i}) $> 4 \times 10^{20}$~cm$^{-2}$ over large regions
(Maybhate et al. 2007).  Thus the tidal bridge between M81--M82 
where Arp's loop is located ($N$(H{\sc i}) $ \sim 5$--$30\times 10^{20}$ cm$^{-2}$; Yun et al. 1994)
appears to support star formation at rates which are
intermediate between the extremely low levels
seen in an object like the Magellanic bridge and
very active tidal tails like those found in, for example, NGC4676
(``The Mice''; de Grijs et al. 2003).  This suggests that the tidal
H{\sc i} streams in the M81 system sit near the low end of a continuum of
levels of star formation in gas-rich tidal matter.

Star formation requires the production of gravitationally bound
molecular clouds. Since tidal streams normally are at most only
mildly gravitationally unstable on large spatial scales, star formation is not expected to
behave in the same way as in the inner disks of galaxies which exist
on the edge of large scale gravitational instability (Kennicutt
1989).  The possibility that collisions between gas clouds could lead
to star formation in low density but gas-rich environments such as
the Magellanic bridge was explored by Christodoulou et al. (1997). In
their model, compression to the densities required to make bound
molecular structures was accomplished through post-shock compression.
This picture gained support as observations improved; e.g. Muller \& 
Parker (2007) make a case for the Magellanic bridge being a prime 
example of star formation resulting from turbulence.

More recent investigations, such as that of Bergin et al. (2004) and Heitsch et al. (2006),
include the effects of turbulence which can enhance density
perturbations more readily leading to molecular regions that can
collapse to make stars in lower velocity cloud collisions.  In these models 
sufficient columns of H{\sc i} must interact with speeds of $\geq$10~km~s$^{-1}$ 
to produce local regions with  $N$(H{\sc i}) $\geq$ 10$^{21}$~cm$^{-2}$ 
that are required to readily shield molecular gas from radiative dissociation. 
The basic idea of triggering star-formation by cloud collision in Arp's
Loop is consistent with the Yun (1999) model for this feature
being a stream crossing zone and with the H{\sc i} observations showing 
comparatively high $N$(H{\sc i}) in a region with a substantial 
$\sim$30~km~s$^{-1}$~kpc$^{-1}$ and complex
velocity field (Rots \&  Shane 1975; Rots 1975; Yun et al. 1994). 

Ongoing star formation
in relatively diffuse structures, such as Arp's Loop, is consistent with expectations for
gas collisions and subsequent turbulence to lead to inefficient
conversion of gas into stars in low density systems such as tidal tails.
On the other hand, the younger star forming sites seem to be relatively normal 
objects with properties similar to the Orion Nebula. Here we find support for the 
expectation that once a gravitationally bound molecular cloud forms in a low 
density environment, its subsequent evolution is largely controlled by factors that
only weakly depend on how the cloud was formed.

\subsection{Connections with Intergalactic HII Regions and Tidal  
Dwarf Galaxies}

The ages and locations of objects containing young stars indicate  
that self-gravitating substructures form within H{\sc i} tidal  
material in a variety of interacting systems.
This tidal matter from interacting galaxies is expected to remain  
bound to one of the galaxies. Much of it will relatively quickly fall  
back into the galaxies, although some
matter may remain detached for as long as a Hubble time. For  
example, according to Hibbard \& Mihos (1995) $\sim$20\% of the tidal  
material of
NGC7252, a well-known merger remnant, is likely to remain in the  
local intergalactic medium. Tidal optical features are also known to  
host
``tidal dwarf galaxies" (TDGs) (e.g. Duc et al. 1997; Weilbacher et  
al. 2000; Knierman, et al. 2003; Bornaud et al. 2004).   The eventual  
fate of the TDGs and the stars and metals that they produce remains  
unclear. For instance, the TDG-candidate found
in the compact group HCG100 (de Mello et al. 2007)  observed with  
GALEX has  young ($<$5~Myr) stellar populations and is located $ 
\sim$100~kpc
away from the main galaxies. These systems therefore have the  
possibility to remain as independent objects over time scales of $\geq$1~Gyr.

The FUV objects in the M81 group are much closer to a large spiral than the HCG100 TDG, as  
they are only  $\sim$17 kpc in projected distance  
from M81 (nucleus) and $\sim$15 kpc from the central region of the  
TDG-candidate, Holmberg~IX.  Besides being located closer to the  
parent galaxies than a long-lived TDG-candidate, the 8 FUV objects  
have much lower UV luminosities than the objects studied
by de Mello et al. (2007) and Neff et al. (2005). As discussed in Sect. 3, their
FUV luminosities are equivalent to $\sim$ 50 late O stars.
The fact that the M81 group is so close and that GALEX is sensitive to very low star-formation levels, 10$^{-3}$ 
M$_{\odot}$ yr$^{-1}$ (Salim et al. 2005), makes it possible to detect the Arp's loop FUV objects. Are we then seeing  
a more diffuse example of star formation, as may be the case in the  
Magellanic bridge, or is the Arp's loop region in an earlier  
evolutionary phase where star formation has not progressed far enough  
to produce a denser region?  Increasing the stellar density is critical for this object to become a TDG (Mizuno et al. 2006).
Better kinematic and baryonic mass  
density information are required to address this question.

It is also possible that the Arp's loop FUV regions are analogs to  
the so-called `intergalactic H{\sc ii}' regions of Mendes de  
Oliveira et al. (2004), Ryan-Weber et al. (2004) and Oosterloo et al.  
(2004).  These objects seem to be similar to the H{\sc ii} regions in  
our Milky Way, but are located in regions with no optical connections  
to nearby galaxies. The fate of these type of objects and their  
importance in galaxy evolution and in the
enrichment of the intergalactic medium are still debatable. These star- 
forming clumps can (1) remain as independent entities and grow to
become TDGs by accreting more gas and forming more stars, (2) make star  
clusters that survive to live in the distant halos of their hosts or  
even become intergalactic objects, or (3) dissolve and not remain  
gravitationally bound yielding only very sparse star streams.

\section{Conclusions}

We have analyzed the UV and optical light within the H{\sc i} bridge between M81 and M82 system and
identified 8 FUV regions in an area known as Arp's loop. Four of these objects were also
detected in H$\alpha$. High resolution ACS images show six of the regions in great detail and allow us to trace their star formation history. 
The synthetic CMD modeling shows the presence of a young population ($<$ 10 Myr) together with an older component with
ages $>$ 1 Gyr. It is possible that these two populations are not related. In any case the young population 
must have formed ``in situ" whereas the old population probably was formed in the M82 and/or M81 galaxies, and removed from their 
disks during a tidal passage $\sim$ 200--300 Myr ago which also triggered star formation in the H{\sc i} debris.

The FUV luminosities of the eight objects we detect are modest and typical of small clusters or associations of O and B stars. 
It is likely that these young stars have condensed 
recently from the tidally stripped gas from M81 and M82. 
The tidal bridge between M81--M82 appears to be intermediate between the very low levels  
of star formation seen in an object like the Magellanic bridge and  
very active tidal tails like those found in interacting galaxies such as the Mice and the Antennae.
Ongoing star formation in relatively diffuse structures, such as Arp's Loop, is consistent with expectations for  
gas collisions and subsequent turbulence to lead to inefficient conversion of gas into stars in low density systems 
such as tidal tails.

The fate of these star-forming clumps is not clear. They can grow to
become TDGs, or make star  
clusters that survive to live in the distant halos of their hosts, or  
even become intergalactic objects. Finally, they may dissolve and not remain  
gravitationally bound, yielding only very sparse star streams. In any case, they will have a composite population of
old and young stars formed in very different environments such as the M81/M82 disks and the tidal bridge. More observations of 
interacting systems similar to the M81 group will help establish the importance of these systems in galaxy evolution and in the
enrichment of the intergalactic medium.  

\acknowledgments
DFdM was funded by STScI grant-44185 and ES was funded by STScI GO grant-10248.07-A. 
JSG thanks the University of Wisconsin for partial support of this research.

GALEX is a NASA Small Explorer, launched in 2003 April. We gratefully acknowledge NASA's support for construction, operation, and science analysis for the GALEX mission, developed in cooperation with the Centre National d'Etudes Spatiales of France and the Korean Ministry of Science and Technology.

This research has made use of the NASA/IPAC Extragalactic Database (NED) which is operated by the Jet Propulsion Laboratory, 
California Institute of Technology, under contract with the National Aeronautics and Space Administration.



{\it Facilities:} \facility{HST (ACS/WFC), \facility{GALEX}, \facility{WIYN}}.



\begin{deluxetable}{cccccccc}
\tabletypesize{\scriptsize}
\tablecaption{FUV sources}
\tablewidth{0pt}
\tablehead{
\colhead{ID} &
\colhead{RA} &
\colhead{Dec} &
\colhead{m$_{V}$} &
\colhead{FUV} &
\colhead{L$_{\rm FUV}$ (erg/s)} &
\colhead{NUV} &
\colhead{L$_{\rm NUV}$ (erg/s)} 
}
\startdata
1+2 & 149.313562  & 69.278637 & 20.25$\pm$ 0.07  &18.17 $\pm$ 0.02 & 1.09E+039 & 19.46 $\pm$ 0.02 & 6.48E+038\\
3   & 149.304548  & 69.273482 &                  &19.77 $\pm$ 0.05 & 2.48E+038 & 21.17 $\pm$ 0.06 & 1.34E+038\\
4   & 149.317475  & 69.273476 &                  &20.03 $\pm$ 0.06 & 1.94E+038 & 21.13 $\pm$ 0.04 & 1.40E+038\\
5   & 149.328273  & 69.273482 & 21.23$\pm$0.05  &20.53 $\pm$ 0.07 & 1.23E+038 & 20.78 $\pm$ 0.05 & 1.92E+038\\
6   & 149.374412  & 69.266872 & 21.52$\pm$0.08  &19.60 $\pm$ 0.04 & 2.90E+038 & 20.45 $\pm$ 0.04 & 2.61E+038\\
7   & 149.393938  & 69.294389 & 22.33$\pm$0.07  &21.33 $\pm$ 0.1  & 5.91E+037 & 22.23 $\pm$ 0.09 & 5.06E+037\\
8   & 149.390473  & 69.296530 & 22.68$\pm$0.09  &21.03 $\pm$ 0.09 & 7.80E+037 & 21.95 $\pm$ 0.08 & 6.52E+037\\
\enddata
\tablecomments
{FUV magnitudes were obtained with SExtractor (Bertin \& Arnouts 1996) Mag$_{-}$auto in the AB system. m$_{V}$ is the magnitude 
within the FUV contour. Galactic extinction corrections were done using Seibert et al. (2005) for FUV and NUV. Objects 1 and 2 are not resolved with 
GALEX and no deblending correction was applied.}
\end{deluxetable}

\clearpage

\end{document}